# Scattering correction based on regularization de-convolution for Cone-Beam CT

Shi-peng Xie[1]
College of Telecommunications and Information Engineering, Nanjing University of Posts and Telecommunications, Nanjing, Jiangsu 210003, China
Rui-ju Yan[1*]
College of Telecommunications and Information Engineering, Nanjing University of Posts and Telecommunications, Nanjing, Jiangsu 210003, China[1]

**Abstract**

In Cone-Beam CT (CBCT) imaging systems, the scattering phenomenon has a significant impact on the reconstructed image and is a long-lasting research topic on CBCT. In this paper, we propose a simple, novel and fast approach for mitigating scatter artifacts and increasing the image contrast in CBCT, belonging to the category of convolution-based method in which the projected data is de-convolved with a convolution kernel. A key step in this method is how to determine the convolution kernel. Compared with existing methods, the estimation of convolution kernel is based on bi-**l1**-**l2**-norm regularization imposed on both the intermediate the known scatter contaminated projection images $g$ and the convolution kernel. Our approach can reduce the scatter artifacts from 12.930 to 2.133.

**1. Introduction**

Computerized Tomography (CT) is a technique for imaging the cross sections of an object using a series of x-ray measurements taken from different angles around the object. An x-ray system with a large-area detector, which is commonly used for cone-beam computed tomography (CBCT), is more susceptible to cupping artifacts generated by scatter and beam hardening. The presence of scatter in the projection images leads to reduced low-contrast sensitivity, artifacts such as dark bands behind dense objects, and slowly varying CT number nonlinearities known as cupping artifacts in the reconstructed 3D images. Beam hardening is also known to lead to image cupping artifacts. X-ray scatter may lead to additional artifacts, which is one of the most challenging problems in CBCT. In the past 20 years, most correction methods that have been developed to reduce the artifacts focus on reducing the scattered photons on the projective image. Various correction methods using software-based [1-7], hardware-based [8-13], or combined hybrid approaches [14-15] have been proposed in the literature. The effect of software-based convolution and de-convolution correction methods depends on whether the chosen conditions which the point spread function and the actual experiment conditions of X-rays in the actual experiment are the same. When the experimental conditions change, we can not guarantee the suppress effect of scatting, therefore the limitation of application is relatively large. Monte Carlo [1-4] simulation is a very effective method of scatter correction, but it is a huge amount of calculation and very time-consuming. Means of hardware-based correction is by choosing experiment equipment that controls and blocks of the photon, which operate more trouble.

[1] Funding information : the National Natural Science Foundation of China (Grant NO. 11547155), National Natural Science Foundation of Jiangsu Province (Grant NO. BK20130883) and the NUPTSF (Grant No. NY213011 and NO. 214026).
Corresponding author email address : Email:1214012235@njupt.edu.cn

In this paper, we propose an improved method to correct for scatter and gain information on the scattering object at the same time. In the proposed method, we probe a convolution-based algorithm for the correction of the image artifacts in projected images. The projected images can be decomposed into two parts: primary projected images and scatter artifacts projected images. To improve the image contrast and reduce the scatter artifacts, we adapted a novel correction method in which we can use a convolution kernel de-convolving with the known scatter contaminated projection images, and obtain the unknown scatter corrected projection images.

## 2. Method

The scatter correction algorithm with the convolution is divided into the following steps:

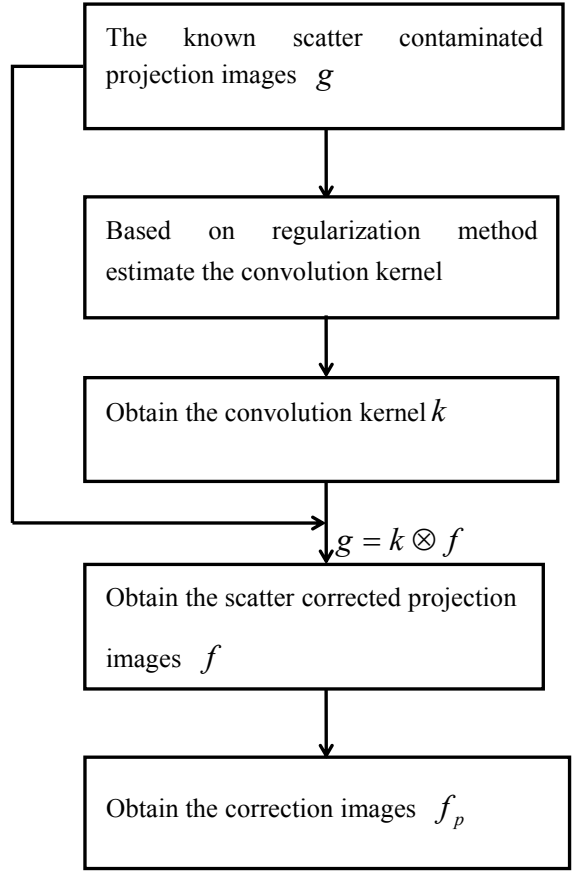

Fig.1 The process of the proposed cupping artifact correction method using de-convolution method.

### 2.1. The algorithm of scatter kernel

We assume that the formation model of the known scatter contaminated projection images is the following:

$$g = k \otimes f + n \qquad (1)$$

Where $g$ denotes the known scatter contaminated projection images, $f$ denotes the unknown scatter corrected projection images, $k$ denotes the convolution kernel or point spread function (PSF) and $\otimes$ denotes convolution operators, and $n$ is assumed to be an additive Gaussian noise. The task of blind convolution is generally separated into two independent stages, i.e., estimation of the convolution kernel $k$ and then a non-blind de-convolution of the unknown scatter corrected projection images $f$ given the found $k$. The contribution in this paper refers to the first stage, which is the core problem of blind de-convolution. It is known that this inverse problem is notoriously ill-posed, and therefore appropriate regularization terms or prior assumptions should be imposed in order to achieve reasonable estimates for the unknown scatter corrected projection images $f$ and the convolution kernel $k$. We should emphasize that the by-product estimated image in the first stage is not necessarily a good reconstruction by itself, and its role is primarily to serve the convolution kernel estimation.

Because of convolution incomplete information, we not only utilize the prior knowledge of the new projection image, but also about the prior knowledge of the constraints convolution kernel estimation in the solution process, that image de-convolution can be modeled as follows minimized title:

$$\min_{f,k} \lambda \|k \otimes f - g\|_2^2 + \sigma \Psi(f) + \beta \varphi(k) \qquad (2)$$

Where $\lambda$、$\sigma$、$\beta$ are positive tuning parameters, $k$ is the convolution kernel, $\Psi(f)$ is the regularization term $\Psi(f) = \frac{\|\nabla f\|_1}{\|\nabla f\|_2}$, $\nabla f$ is the gradient of the estimated primary images; $\|\bullet\|_1$、$\|\bullet\|_2$ represent $l_1$ norm and $l_2$ norm of images, respectively.

Our kernel estimation is performed on the high frequencies of the image. Given the known scatter contaminated projection images g, we use discrete filters $\nabla_x = [1;-1]$ and $\nabla_y = [1;-1]^T$ to generate a high-frequency version y = $[\nabla_x g, \nabla_y g]$. The energy function for spatially invariant convolving kernel is:

$$\min_{x,k} \lambda \|k \otimes x - y\|_2^2 + \frac{\|x\|_1}{\|x\|_2} + \beta \|k\|_1 \qquad (3)$$

It is subject to the restrictions that $k \geq 0, \sum k_i = 1$; Here $x$ is the images in the high-frequency space, $k$ is the unknown convolving kernel ($k_i$ are individual elements) and is the 2D convolution operator.

Eq.3 consists of 3 terms. The first term is the likelihood that takes into account the formation model Eq.1. The second term is the regularization on $x$ which increase scale-invariant scarcity in the reconstruction process. The third term is in order to reduce noise in the kernel, we add regularization on $k$. The scalar weights $\lambda$ and $\beta$ controls the relative strength of the kernel and images regularization terms.

Eq.3 is highly non-convex. To optimizing such a problem, the standard method initializes on $x$ and $k$ at first, and then alternate between $x$ and $k$ updates [16].

### 2.1.1. $x$ Update

The $x$ sub-problem is given by the Eq.4:

$$\min_{x} \lambda \|x \otimes k - y\|_2^2 + \frac{\|x\|_1}{\|x\|_2} \tag{4}$$

Eq.4 is non-convex because of the presence of the regularization term $\frac{\|x\|_1}{\|x\|_2}$. However, if one fixes the denominator of the regularization from the previous iterate, the issue becomes a convex $l_1$-regularized problem. Fast algorithms to solve $l_1$-regularized problems are well known in the compressed sensing literature [17-18]. One such algorithm is fast the iterative shrinkage-threshold algorithm (FISTA). FISTA is a fast method to solve general linear inverse problems of the form. The model of general linear problem is following:

$$\min_{x} \lambda \|Kx - y\|_2^2 + \|x\|_1 \tag{5}$$

In our application $K$ is the convolving kernel matrix. The operator S is a vector on the FISTA algorithm.

$$S_\alpha(x)_i = \max(|x_i - \alpha|, 0) sign(x_i) \qquad (6)$$

FISTA is very simple and fast, involving only multiplications of the matrix $k$ with vector $x$, followed by the component-wise shrinkage operation.

### 2.1.2. $k$ Update

After updating x, we update k. The kernel update sub-problem is given by:

$$\min_k \lambda \|x \otimes k - y\|_2^2 + \beta \|k\|_1 \qquad (7)$$

It is subject to the restrictions that $k > 0, \sum_i k_i = 1$. We use unconstrained iterative re-weighted least squares (IRLS) [19] followed by a projection of the resulting k onto the restrictions (setting negative elements to 0, and re-normalizing).

### 2.2. The algorithm of scatter correction

Once the kernel $k$ has been estimated, we can use a variety of de-convolution methods to recover the unknown scatter corrected projection images $f$ from g. The simplest is Richardson-Lucy (RL). The disadvantage of RL is that this method is sensitive to a wrong kernel estimate, which results in ringing artifacts in $f$. Therefore, we choose to use the de-convolution method from [20], since it is fast and robust to small kernel errors. This algorithm uses TV regularization, the regularization model:

$$\min_f \|f \otimes k - g\|_2^2 + \lambda TV(f) \qquad (8)$$

Where $\lambda$ is the parameter of regularization $TV(f) = \sum_i \sqrt{(\Delta_i^h)^2 + (\Delta_i^v)^2}$ ; $\Delta_i^h$、$\Delta_i^v$ represent the horizontal of the first-order differential operator and the vertical of the first-order differential operator at the pixel $i$, respectively. $\Delta_i^h = f_i - f_j$、$\Delta_i^v = f_i - f_k$  $f_j, f_k$ represent first-order neighborhood pixel gray values of the left and top of $f_i$.

Because of having the scatter corrected projection images, we can reconstruct images basing on FDK [21]. As show the following two figures, there are the reconstruction image and the scatter correction reconstruction image by their projection data. It is clear that these artifacts are significantly suppressed when the proposed method is used with the de-convolution. From the following two Figs, we found that the shading artifacts were successfully reduced and the missing details can be clearly seen in Fig.3.

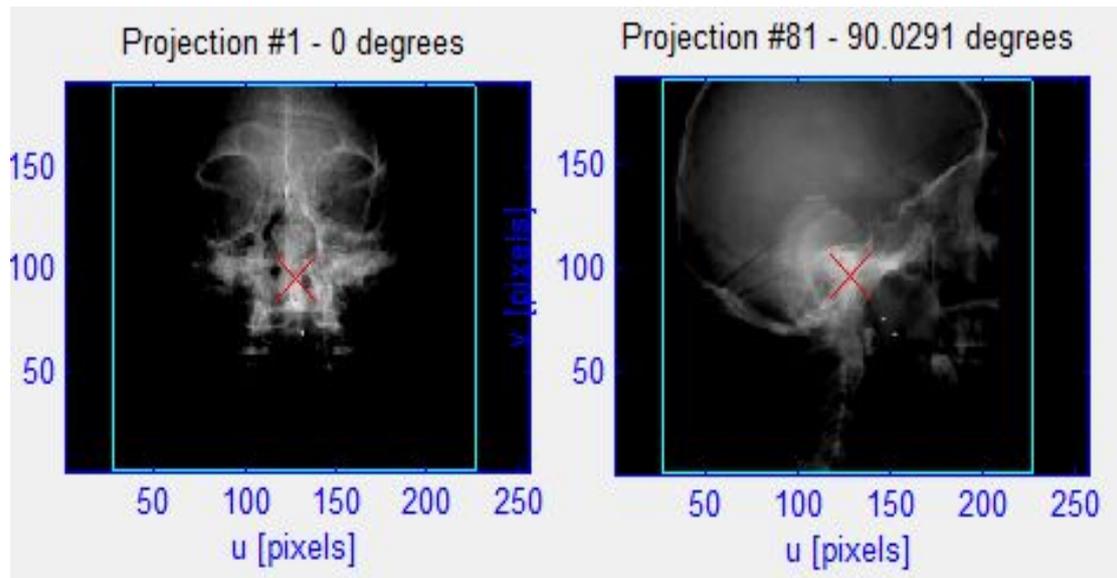

Fig.2: the original projection image

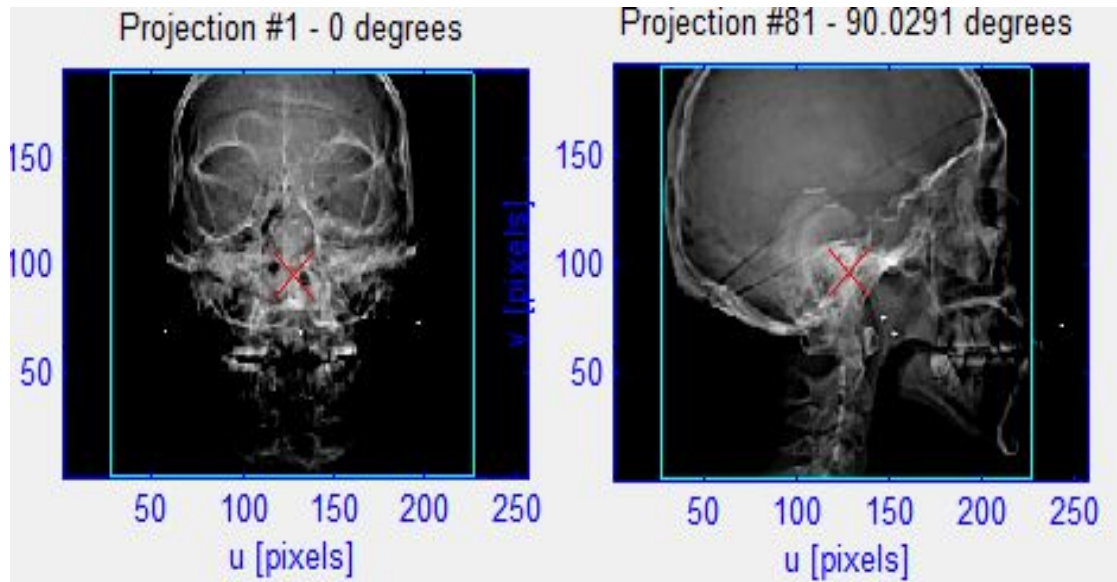

Fig.3: the new projection image

**2.3. Evaluation of scatter correction**

The quantitative image quality parameters were analyzed according to Ref. 12. For the contrast-to-noise ratio (CNR), the regions of interest (ROIs) with the same radius of the contrast-and-resolution phantom were analyzed. The measured (denoted by the subscript M) mean values of the ROIs are $(u_{M,1}, u_{M,2})$, and the standard deviations are $(\sigma_{M,1}, \sigma_{M,2})$. The voxel noise $(\sigma_M)$ is calculated as $\sigma_M = (\sigma_{M,1} + \sigma_{M,2})/2$, and the CNR is calculated as $CNR = |u_{M,1} - u_{M,2}|/\sigma_M$.

The magnitude of cupping $\tau_{cup} = 100(u_{M,edge} - u_{M,center})/u_{M,edge}$, is extracted in terms of the pixel values at the center $u_{M,center}$ and edge $u_{M,edge}$ of the phantom.

The error of the CT number in the ROI is calculated as the root mean square error (RMSE), which is defined as:

$$E_{RMSE} = \sqrt{mean\left[(I_{ij} - \bar{I})^2\right]} \tag{8}$$

Where $I_{ij}$ are the $i$ th and $j$ th elements of the ROI and $\bar{I}$ is the mean reconstructed value inside the ROI.

## 3. Results

**The tabletop CBCT systems**

The system parameters of the tabletop CBCT systems used in this paper are summarized in Table I.

Table I : Imaging parameters of the physical experiments.

| Parameters | System values |
| --- | --- |
| Number of projections | 320 |
| Source-Axis-Distance | 100 cm |
| Source-Detector-Distance | 150 cm |
| Detector size | $254 \times 190 \ mm^2$ |
| Detector pixel array | $192 \times 256$ |
| Anti-scatter grid | No |
| Bow-tie filter | No |
| Reconstructive method | FDK |

**Experiments on Phantom I**:

The phantom is the projected data which is de-convolved with a convolution kernel. Fig.5 and Fig.6 shows the reconstructed images. As in Fig.4(a) and Fig.6(c), Because of the scatter, the image distortion (cupping etc.)is obvious in the reconstructed images without scatter correction. The scatter artifacts are greatly reduced in the images using the proposed method, which can be seen in the Fig.4(b) and Fig.6(e).

The 1D horizontal profile with and without scatter signals can be seen in Fig.5and Fig.7. As in the figure, a reduction in the cupping artifacts when the image is corrected for scatter radiation is observed.

For a quantitative analysis of the reconstructed image, we measured the image contrast and the magnitude of cupping in Fig.4 (a) and Fig.6(c). The results can be found in Table II and Table III. Our approach can increase the image CNR from 5.61 to 7.35 and reduce the artifact from 12.930 % to 2.133 %. As show in Table II and Table III, these artifacts are significantly suppressed when the proposed method is used with convolution

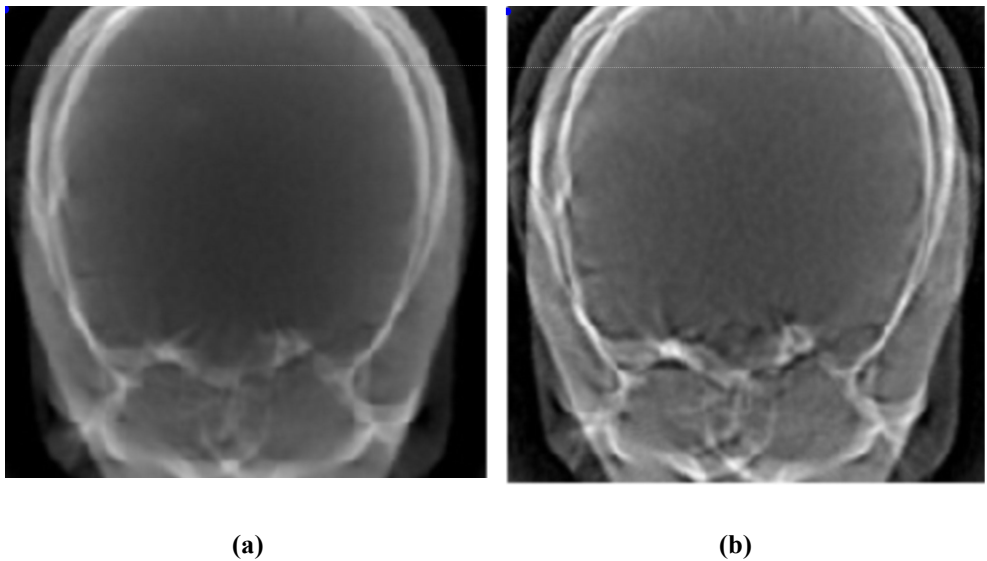

(a)          (b)

Fig.4: Image reconstructions of the phantom. (a) CBCT without scatter correction; (b) scatter correction using the de-convolution method.

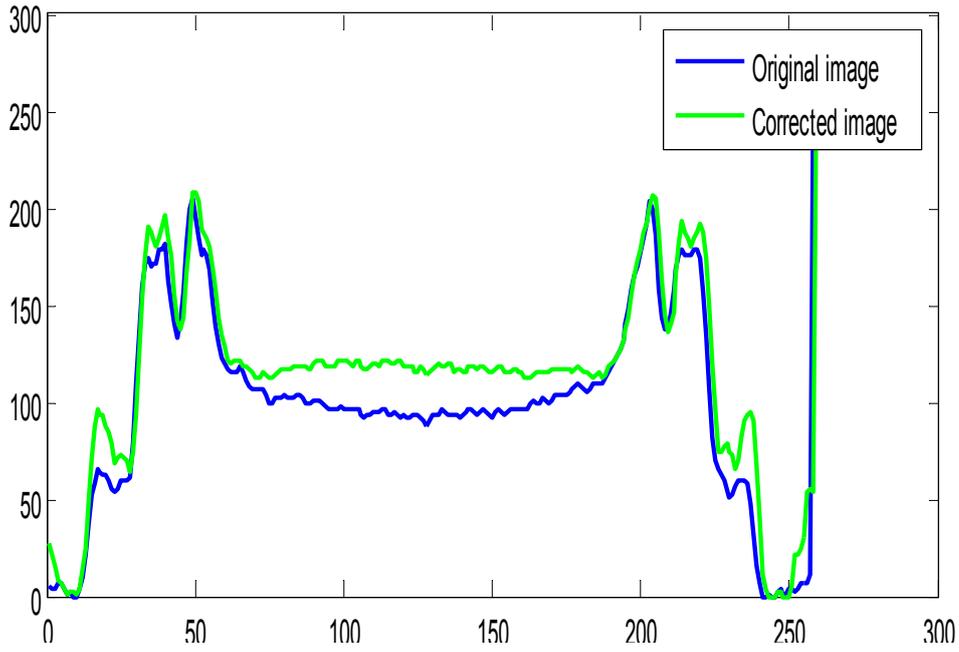

Fig.5: The 1D horizontal profile of the measured and estimated scatter signals on the phantom: The column is the profiles of the projection image which is at row 26 in Fig. 4.

Table II: Quantitative analysis of Fig4 (a) (b)

| Modality | $E_{RMSE}$ (edge) | $E_{RMSE}$ (center) | $\tau_{cup}$ in [%] | CNR |
|---|---|---|---|---|
| CB_NONE | 6.171 | 8.895 | 9.435 | 5.61 |
| CBCT TC | 8.365 | 10.324 | 3.422 | 7.35 |

Note: CBCT without scatter correction (CBCT NONE), scatter correction using the convolution method (CBCT TC) and $\tau_{cup}$ are taken from the phantom.

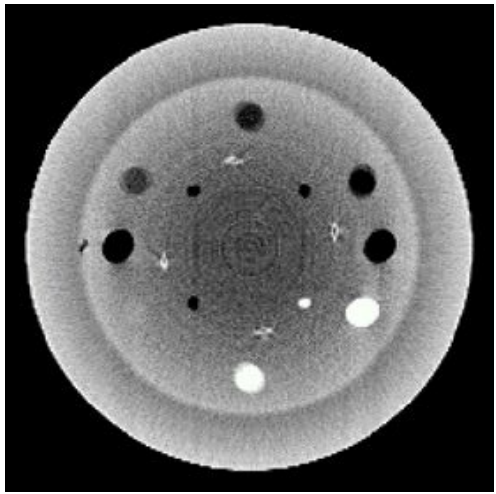 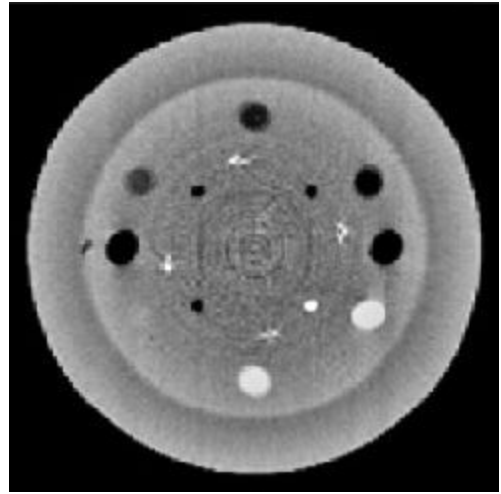

(c)  (d)

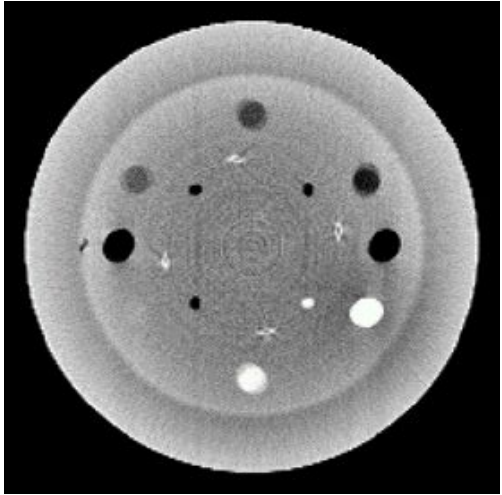

(e)

Fig.6: Image reconstructions of the phantom. (c) CBCT without scatter correction; (d) scatter correcting using other method (e) scatter correction using the de-convolution method

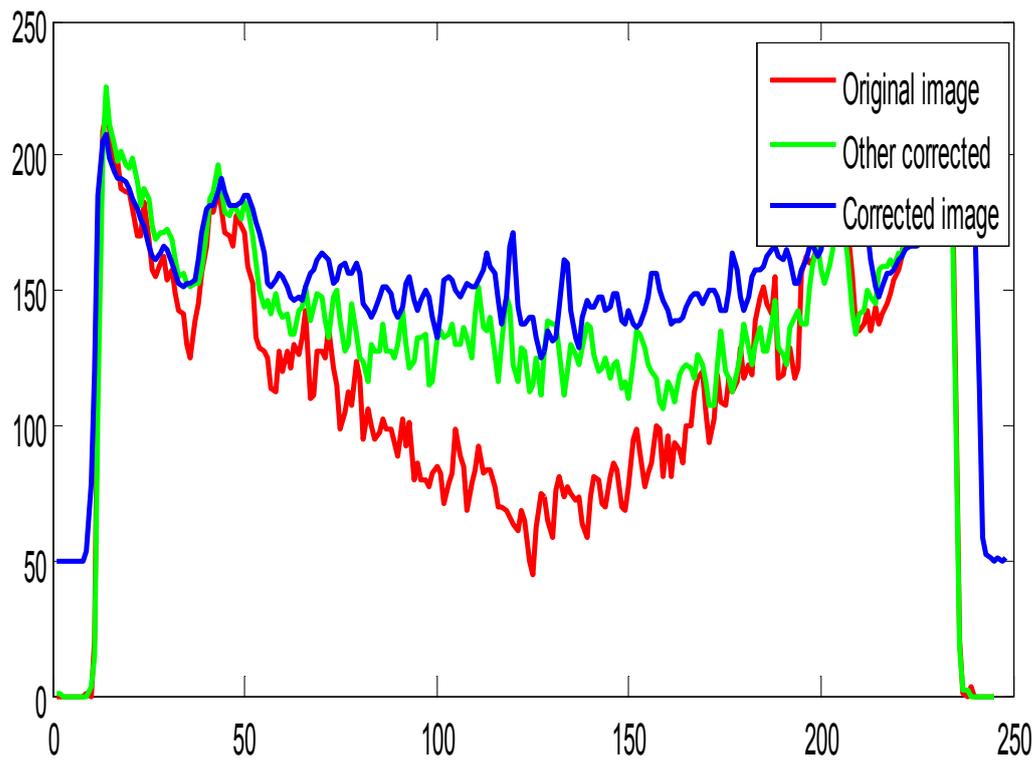

Fig.7: The 1D horizontal profile of the measured and estimated scatter signals on the phantom: The column is the profiles of the projection image which is at row 110 in Fig. 6.

Table III :Quantitative analysis of Fig6(c)(d)(e)

| Modality | $E_{RMSE}$ **(edge)** | $E_{RMSE}$ (center) | $\tau_{cup}$ in [%] |
|---|---|---|---|
| CB_NONE | 3.171 | 2.655 | 12.930 |
| CBCT OTC | 3.916 | 2.870 | 4.544 |
| CBCT TC | 4.364 | 3.390 | 2.133 |

Note:CBCT without scatter correction (CBCT NONE), scatter correction using the convolution method (CBCT TC) ,scatter correcting using other method (CBCT OTC)and $\tau_{cup}$ are taken from the phantom.

**Experiments on Phantom II**:

The phantom is the projected data which is de-convolved with a convolution kernel. Fig.7 shows the reconstructed images. As in Fig.7(a) (c), Because of the scatter, the image distortion (cupping etc.) is obvious in the reconstructed images without scatter correction. The scatter artifacts are greatly reduced in the images using the proposed method, which can be seen in Fig.7(b) (d)

The 1D horizontal profile with and without scatter signals can be seen in Fig.8. As in the figure, a reduction in the cupping artifacts when the image is corrected for scatter radiation is observed.

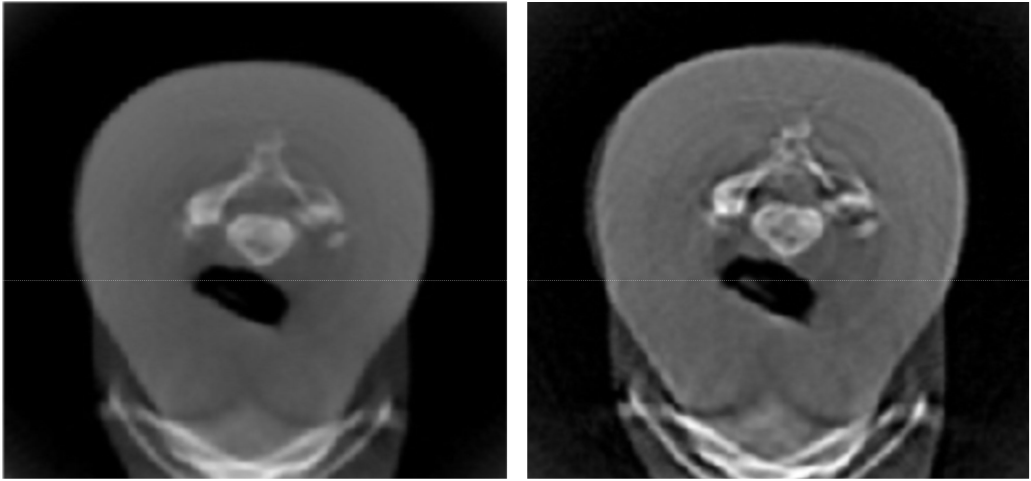

(a)                            (b)

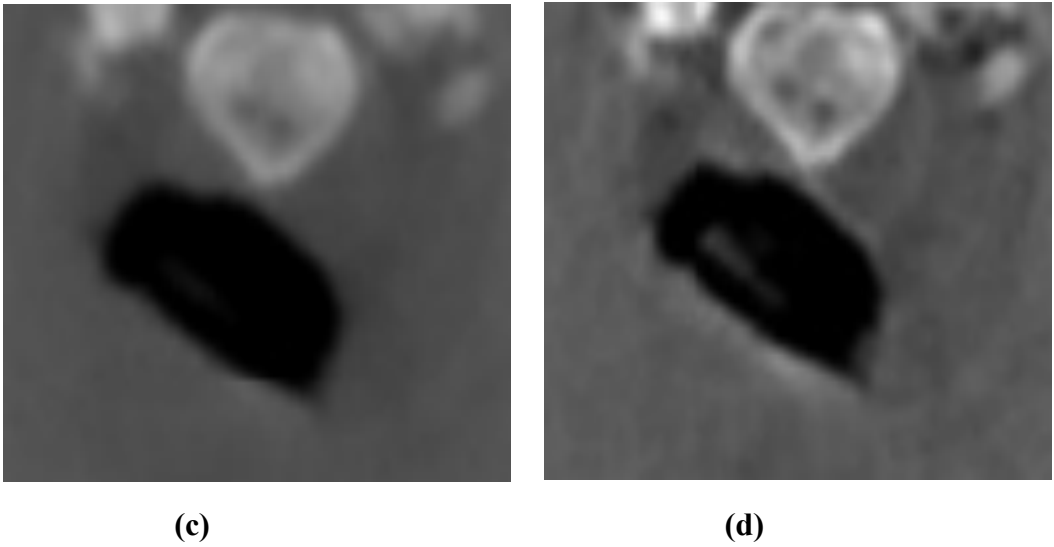

**(c)** **(d)**

Fig.8: Image reconstructions of the phantom. (a)(b) CBCT without scatter correction; (b) (d) Scatter correction using the de-convolution method.

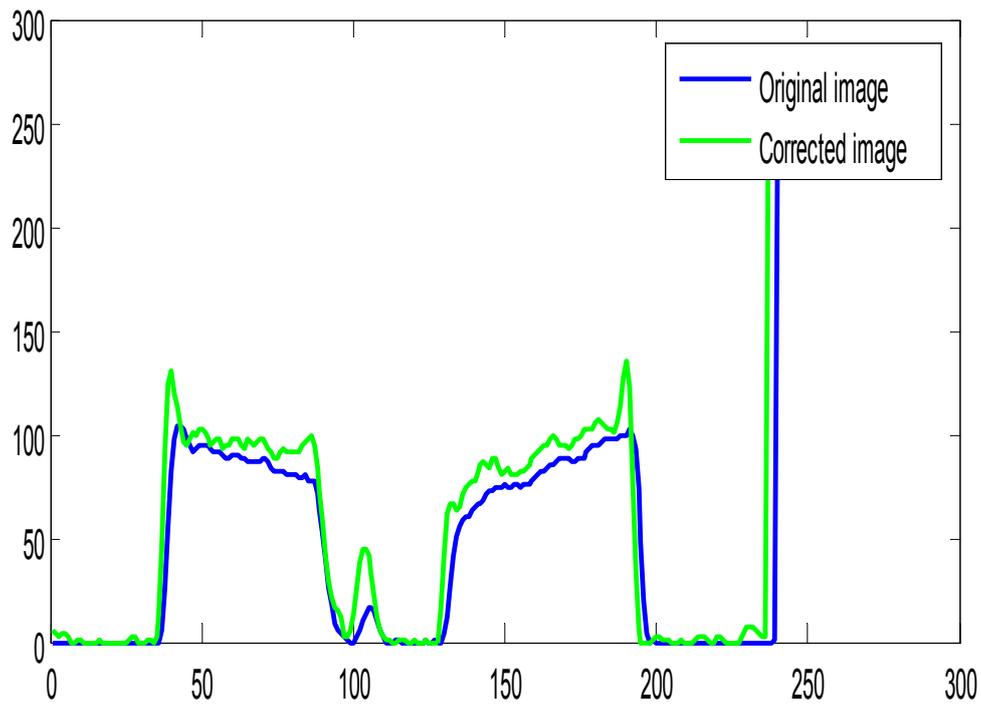

Fig.9: The 1D horizontal profile of the measured and estimated scatter signals on the phantom: The column is the profiles of the projection image which is at row 127 in Fig. 8.

## 4. Discussion

In this paper, a new convolution-based method for the correction of cupping artifacts in CBCT images was developed. By using this correction technique, we can improve the image CNR. The method is completely different from the previous convolution method; the advantages of this de-convolution correction method are as follows:

First, this method is very practical for clinical use in shading correction where artifacts are low-frequency and continuous, including cupping artifacts resulting from scatter contamination and large-area shading artifacts from beam hardening. As shown in Figs.4-7 and Tables.II-III, we can find the proposed method reduced the magnitude of scatter artifacts, and increased the image contrast, and improve the RMSE at the image edges and center. The Figs.8-9 show that the uncorrected images show shading artifacts and part of the information was covered. After scatter correction, we can find the covering parts. Second, this method is computationally efficient. All steps for implementation are standard signal and image processing techniques.

In addition, another scatter correction methods basing on convolution were found in the literature [23-24]. We can find, in the solution process of the method, they need to do a lot of simulation experiments so as to determine convolution kernel, and then calculate the scattering estimates. There is a general drawback of scatter correction methods in which the estimated scatter signals are subtracted from the raw projection data. This subtraction process often amplifies the noise, and image de-noising techniques can be employed to reduce noise levels.

## 5. Conclusion

An effective de-convolution algorithm for scatter correction in CBCT imaging is proposed. The result shows that our algorithm produces substantial image quality improvements. Firstly, the proposed method increases the image CNR of CBCT. Secondly, the magnitude of scatter artifacts is also descended. At the same time our method is practical and attractive as a general solution to CT shading correction and without the loss of real-time imaging capabilities, which makes treatment planning using CBCT images a viable option.

**Acknowledgments**

This work was supported by the National Natural Science Foundation of China (Grant NO. 11547155), National Natural Science Foundation of Jiangsu Province (Grant NO. BK20130883) and the NUPTSF (Grant No. NY213011 and NO. 214026).